\documentclass[twocolumn,preprintnumbers,amsmath,amssymb,prb,floatfix]{revtex4-2} 
\usepackage{graphicx}
\usepackage{dcolumn}
\usepackage{bm}
\usepackage{float}
\usepackage{color}
\usepackage[dvipsnames]{xcolor}
\usepackage{hyperref}
\usepackage{graphicx}
\usepackage{braket}
\usepackage{appendix}
\usepackage{chemmacros}
\usepackage{multirow}
\usepackage{natbib}
\usepackage{amsmath}
\hypersetup{colorlinks=true, citecolor=blue, urlcolor=blue, linkcolor=blue}
\usepackage{soul}

\usepackage{multirow}

\begin{document}



\title{
Pressure-correlation phase diagram unraveling quantum phase transitions in altermagnetic $\epsilon$-FeOOH and realization of piezo-anomalous Hall effect
 }  

\author{Arijit Mandal}\altaffiliation[arijitmandal1997@gmail.com]{}\affiliation{Condensed Matter Theory and Computational Lab, Department of Physics, IIT Madras, Chennai-600036, India}
\affiliation{Center for Atomistic Modelling and Materials Design, IIT Madras, Chennai-600036, India}
\author{Nayana Devaraj}\altaffiliation[]{}\affiliation{Condensed Matter Theory and Computational Lab, Department of Physics, IIT Madras, Chennai-600036, India}
\affiliation{Center for Atomistic Modelling and Materials Design, IIT Madras, Chennai-600036, India}
\author{Birabar Ranjit Kumar Nanda}\altaffiliation[nandab@iitm.ac.in]{}\affiliation{Condensed Matter Theory and Computational Lab, Department of Physics, IIT Madras, Chennai-600036, India}
\affiliation{Center for Atomistic Modelling and Materials Design, IIT Madras, Chennai-600036, India}

\begin{abstract}
We establish a pressure-correlation phase diagram that yields multiple pressure-tunable quantum phase transitions in iron oxyhydroxide ($\epsilon$-FeOOH) across three critical pressures, $P_{c1}>P_{c2}>P_{c3}$. The transitions arise from the competition among electron-correlation, crystal-field, and Hund’s coupling. First-principles calculations reveal that the earlier reported high-spin (HS) to low-spin (LS) crossover at correlation-sensitive $P_{c1}$ is accompanied by a transition from charge-transfer to Mott insulating (CTI-MI) phase, and a change in the spin-quantization axis (SQA) from $\hat{y}$ to $\hat{x}$. The HS-LS crossover and CTI-MI transition is driven by the simultaneous increase in crystal-field splitting and decrease in Hund’s coupling. The critical pressure $P_{c2}$ governing the structural transition from centrosymmetric $Pnnm$ phase to noncentrosymmetric $P2_1nm$ phase is found to be independent of electron-correlation. When correlation is strong, a third critical pressure ($P_{c3}$) emerges in the lower-pressure regime of the phase diagram. Validating the recent experimental observation, $P_{c3}$ is responsible for changing the SQA from $\hat{z}$ to $\hat{y}$. By exhibiting altermagnetism in both $P2_1nm$ and $Pnnm$ phases and possessing finite anomalous Hall conductivity (AHC) only for SQA along $\hat{x}$ and $\hat{y}$,  $\epsilon$-FeOOH enables a unique piezo-anomalous Hall effect where one can switch on the AHC at a lower pressure and change its direction at a higher pressure.
\end{abstract}

\maketitle

Correlated systems, due to their rich electronic and magnetic phase diagram, have always remained at the forefront for exploring emerging quantum phases and insights in condensed matter physics \cite{RMP_tokura, Paschen2021, Hujiben_nat_mat, science_nagaosa, Schaffer_2016, kim_science, nanda_cro_cmo, priyo}. Oxide minerals are often considered the appropriate class of materials for investigating these phases and their transitions, as they can withstand high pressures and temperatures. In recent years, due to the existence of large interstitial space in the form of tunnels and vander Waals separations, encouraging quantum confinement and strong octahedral crystal fields, the iron oxyhydroxide polymorphs \cite{theory_Rossitza, alpha_FeOOH1, alpha_FeOOH2, FeOOH-JACS, beta_FeOOH1, beta_FeOOH2, gamma_FeOOH1, gamma_FeOOH2, gamma_FeOOH3}, namely goethite, akaganeite, and lepidocrocite, have gained renewed attention in the context of magnetism, structural breakdown and transitions, and desirable electrochemical activities. 

\begin{figure}[hbt!]
    \centering
   \includegraphics[width=1\linewidth]{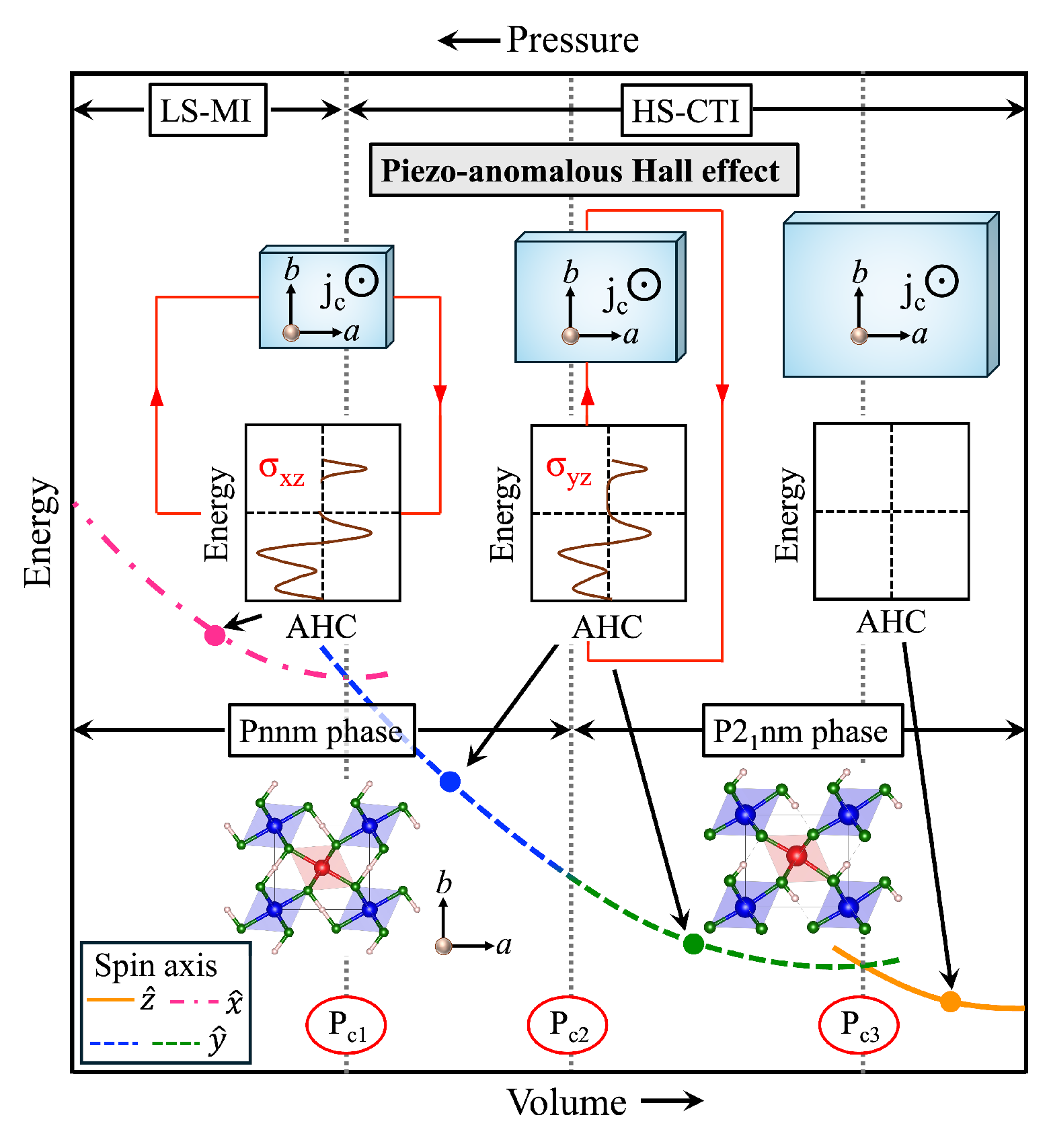}
    \caption{\textbf{Schematic illustration of pressure-driven structural and quantum phase transitions.} The $\epsilon$-FeOOH undergoes a structural transition from $P2_1nm$ to $Pnnm$ at a critical pressure ($P_{c2}$) and spin crossover from HS-CTI to LS-MI state at a different critical pressure ($P_{c1}$). The distinction is shown through the free energy curves.  Furthermore, at $P_{c1}$ and $P_{c3}$ the spin quantization changes from $\hat{x}$ to $\hat{y}$ to $\hat{z}$, respectively. The former two exhibit a finite anomalous Hall conductivity (AHC). The change in spin quantization from $\hat{z}$ to $\hat{y}$ axis switches on the AHC, while the change of the same from $\hat{y}$ to $\hat{x}$ axis rotates the direction of AHC.}
    \label{toc}
\end{figure}

Particularly, the quenchable goethite, $\epsilon$-FeOOH, which can endure phenomenally high pressure (around 100 GPa) and can shrink volume by $50\%$, has emerged as a reference system as it demonstrates switchable spin states, tunable spin anisotropy, and structural transitions \cite{theory_Insixiengmay, thompson_2020, gleason_2013, FeOOH-JACS}. 

A brief review of the literature provides the following information about the critical pressures across which the phase transitions occur. From the experimental perspective, the structure has a second-order transition ($P2_1nm \rightarrow Pnnm$ phase) around 18 GPa and a first-order phase transition around 45 GPa \cite{gleason_2013, thompson_2020, FeOOH-JACS}. The theoretical studies propose the structural transition pressure to be in the range from 37-45 GPa \cite{gleason_2013, theory_Insixiengmay}. Concerning the high spin (HS) to low spin (LS) crossover, both from a theoretical and experimental perspective, there is a lack of clarity, and it lies somewhere in between 45 and 70 GPa \cite{gleason_2013, thompson_2020, theory_Insixiengmay}. In addition, a recent experimental studies by Ikeda \textit{et al.} propose a possible change in the spin quantization axis in the system at 8 GPa \cite{FeOOH-JACS}. In the absence of a complete electronic and magnetic phase diagram in the space spanned by pressure and correlation, the understanding of the underlying mechanism driving these transitions is missing. Also, with several competing interactions at play, there can be other missing quantum transitions in this class of minerals.

Furthermore, with the recent discovery of altermagnetism, identified with non-relativistic spin-splitting of the bands of a real space antiferromagnet \cite{libor1, libor2}, where fundamental understanding is still emerging, and non-trivial quantum transports are being envisaged \cite{song2025altermagnets,bai2024altermagnetism,10.1039/d6cs00793g, arijit_NiS, Nayana, Arindom}, the FeOOH mineral class, with its spin anisotropy and pressure tunable symmetries, can serve as a promising family to investigate. 

In this work, we have carried out electronic and magnetic structure calculations, and analyzed the results from the perspective of a strong correlation effect  (i) to establish a complete phase diagram, (ii) to present the underlying mechanism responsible for the reported quantum phase transitions, and (iii) to unravel yet to be reported phase transitions and phases including altermagnetism. 

The critical findings from this work, summarized in Fig. \ref{toc}, are as follows. There are three correlation-tunable critical pressures across which the phase transitions occur. Validating the majority of the experimental observation, the structural transition from $P2_1nm$ (non-centrosymmetric) to $Pnnm$ (centrosymmetric), at a critical pressure $P_{c2}$, occurs while the system is in the HS phase (intermediate pressure), and the crossover to the LS state occurs at a higher critical pressure ($P_{c1}$). Interestingly, $P_{c2}$ is almost independent of the correlation effect, while $P_{c1}$ is sensitive to it. Consequently, concurrent with the HS-LS crossover, from the electronic structure perspective, the system changes from a charge transfer insulator (CTI) to a Mott insulator (MI) type, which is an outcome of competition between Hund's coupling and crystal field. Symmetry compatibility ensures altermagnetism in both the structural phases, and therefore, it makes this system unique in achieving pressure switchable non-centrosymmetric to centrosymmetric altermagnet. The spin quantization axis alters its direction from $\hat{z}$ to $\hat{y}$ to $\hat{x}$ at two critical pressures $P_{c3}$ and $P_{c1}$, respectively; one at the HS regime and one at the crossover of HS to LS, which enables the system to exhibit piezo-anomalous Hall effect, a phenomenon yet to be reported. 

\begin{figure*}[hbt!]
    \centering
   \includegraphics[width=1\linewidth]{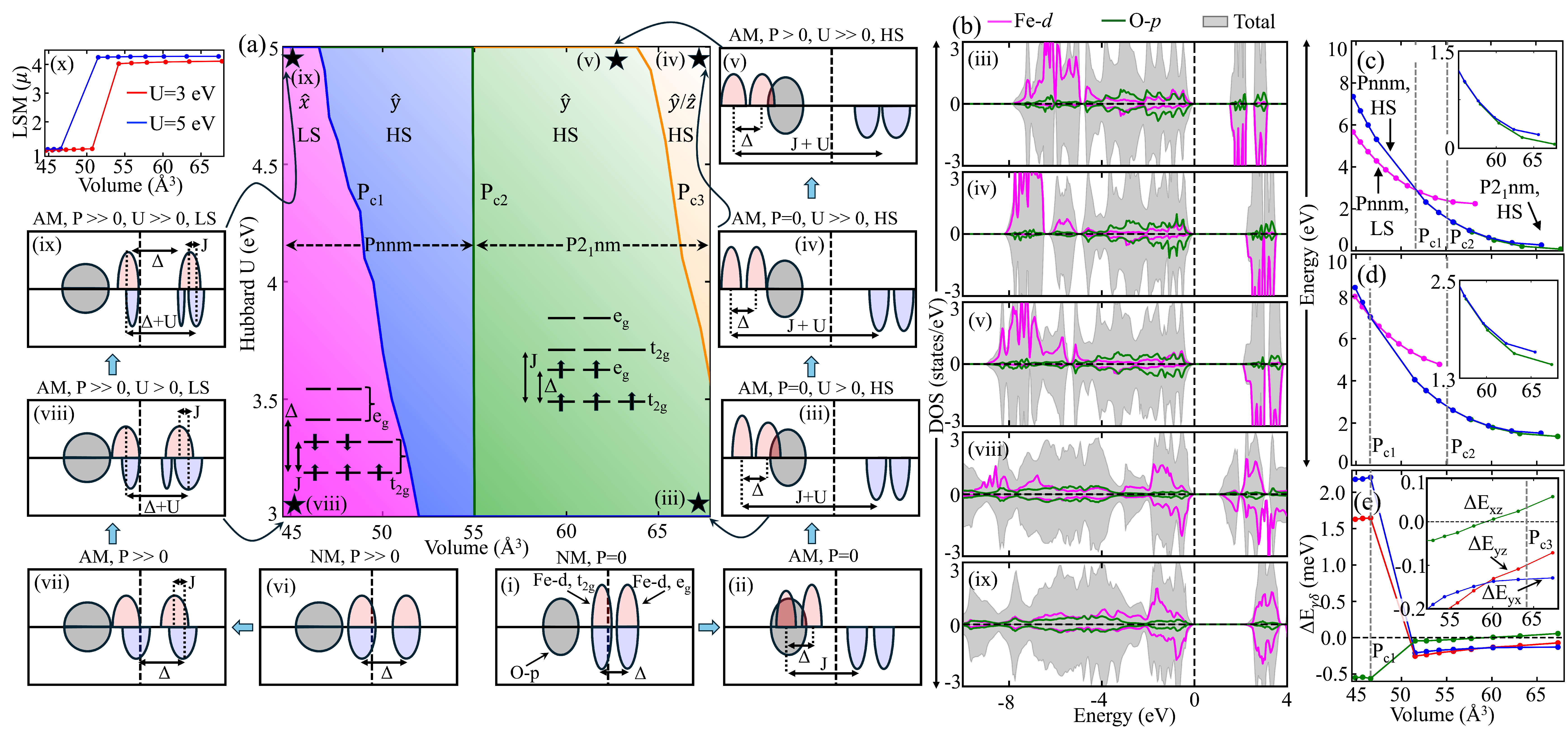}
    \caption{\textbf{Correlated electron physics: Structural and magnetic phase diagram for $\epsilon$-FeOOH in the space spanned by onsite Coulomb repulsion U and volume as obtained from First-principles calculations.} 
    (a) The phase diagram at the center displaying the $P2_1nm$ to $Pnnm$ structural phase transition at moderate compression, the  HS to LS transition at higher compression, and the flipping of the spin quantization $\hat{y}/\hat{z}\ \rightarrow \ \hat{y} \ \rightarrow \ \hat{x}$ with increase in compression.  The strong correlation affects the critical compression for HS $\rightarrow$ LS much more than that of the $P2_1nm \rightarrow Pnnm$ structural transition. The intensity gradient reflects the extent of delocalization of the states. The higher the intensity, the higher the delocalization. The extent of delocalization is governed by both compression and onsite Coulomb repulsion U. In the inset of the phase diagram, the relative positioning of the octahedral crystal field split ($\Delta$) $t_{2g}$ and $e_g$ energy labels for both the spin channels, controlled by Hund's coupling $J$, is schematically illustrated. Our density functional theory (DFT) calculation suggests that at high compression, $\Delta$ is larger than $J$ to favor an LS state. At low compression, $\Delta$ is smaller than $J$ to favor a high spin state. The strength of the LSM as a function of compression for two specific U values is shown on the top left. The schematics surrounding the phase diagram summarize the evolution of electronic structure as a function of compression and strong correlation. (b) The spin-resolved total and orbital projected DOS for the points marked by a star in the phase diagram. The schematic illustrations are based on these specific cases, along with a few others (not shown here). These schematics, in addition to demonstrating the effect of crystal field splitting and spin polarization, explain the mechanism that stabilizes the Mott insulating picture for the LS state and the charge transfer insulating picture for the HS state. (c)-(d) The DFT+U obtained relative energies of various phases (with respect to the energy of the $P2_1nm$ phase at 0 GPa and $U = 3$ eV), such as $P2_1nm$ HS (green), $Pnnm$ HS (blue), and $Pnnm$ LS (magenta), for $U=3$ and $U=5$ eV, respectively. (e) The DFT+U obtained magnetic anisotropy energy $(\Delta E_{\gamma \delta} = E_{\hat{\gamma}} - E_{\hat{\delta}})$ as a function of compression. The green curve ensures the $\hat{x}$ axis as spin quantization axis in the LS state, while the blue and red curves together indicate the $\hat{y}$ axis as spin quantization axis in the HS state. Furthermore, the red curve on the right of $P_{c3}$ indicates the near degeneracy of $\hat{y}$ and $\hat{z}$ axes for the spin quantization axis in the zero or low pressure regime. Therefore, $P_{c1}$ and $P_{c3}$ work as a switch to induce the direction-dependent piezo-anomalous Hall effect in this altermagnetic material.
    }
    \label{phase}
\end{figure*}

\section*{Results}
\noindent
\textbf{Pressure-correlation phase diagram}

\noindent
The phase diagram is computed by performing DFT+U calculations on relaxed structures by varying the onsite Coulomb repulsion ($U$) and the volume of the $\epsilon$-FeOOH unit cell. The structural and computational details, as well as the relation between pressure and volume, are provided in the appendix. The phase diagram is shown in Fig. \ref{phase}(a). The three distinct critical pressure curves, $P_{c1}(U) > P_{c2}(U) > P_{c3}(U)$ identifying the multiple phase transitions are shown. 

We will first discuss the case of $P_{c1}(U)$, which exhibits the HS-LS crossover as well as the CTI to MI transition by mapping our DFT results to the following multiorbital Hubbard Hamiltonian \cite{amit_sro, mandal_PRM, chain_doped}:
\begin{eqnarray}
    \mathcal{H} &=& \sum_{i,\mu} \epsilon_{i\mu} c_{i\mu}^{\dagger}c_{i\mu} + \sum_{i,j,\mu,\nu} t_{i\mu j \nu} c_{i\mu}^{\dagger} c_{j\nu} + \text{H.c.} \nonumber \\
    && + \mathcal{H}_{\text{corr}} + \mathcal{H}_{\text{mag}} \nonumber \\
    \mathcal{H}_{\text{corr}} &=& U\sum_{i,\mu} n_{i\mu \uparrow} n_{i\mu \downarrow} + \left( U'-\frac{J}{2} \right)\sum_{i,\mu < \nu} n_{i\mu} n_{i\nu} \nonumber \\
    \mathcal{H}_{mag} &=& -2J\sum_{i,\mu< \nu} S_{i\mu}^z \cdot S_{i\nu}^z
    \label{Hubbard}
\end{eqnarray}
where $U' = U - 2J$ and $n_{i\mu} = n_{i\mu \uparrow} + n_{i\mu \downarrow}$. Here, the first and second terms represent the onsite energy and electron hopping, respectively. The third and fourth terms (i.e., $\mathcal{H}_{\text{corr}}$ and $\mathcal{H}_{\text{mag}}$) consist of Hubbard interactions among electrons in the same orbital, different orbitals, and Hund's coupling energy, respectively. The mean-field expansion of the Hubbard terms, involving orbital and spin space occupancies, is provided in the appendix.

With $3+$ charge state of Fe in $\epsilon$-FeOOH, the $3d^54s^0$ electronic configuration, under the influence of octahedral crystal field $\Delta$ and Hund's coupling $J$, governs the electronic and magnetic structure of the system.
$J$ is primarily an atomistic phenomenon, while $\Delta$, which gets incorporated in the first term of Eq. \ref{Hubbard}, is driven by the volume and shape of the octahedron. In addition to this, the electron hopping $t$ and repulsion $U$, both of which depend on the atomistic characteristics as well as the crystal structure, complicate the electron filling scenarios \cite{priyadarshini_ppr}. 
The strength of $\Delta$ that splits the triply degenerate $t_{2g}$ manifold and doubly degenerate $e_g$ manifold increases with compression. In a nonmagnetic state, the splitting leads to a five-sixths occupied $t_{2g}$ state and an unoccupied $e_g$ state, and they are well-separated in the equilibrium structure as shown in the schematic Fig. \ref{phase}(a-i). The large DOS at the Fermi level ($E_F$) compels the system to lower its energy through spin polarization with the help of Hund's exchange splitting $J$. As the system is half-filled, a strong $J$ ($> \Delta$), driven by the last term of Eq. \ref{Hubbard}, helps the Fe to achieve the HS state with $S=5/2$. The Fe-O-Fe superexchange stabilizes the antiferromagnetic state, and it results in a Slater insulator as demonstrated in Fig. \ref{phase}(a-ii).

Upon application of Coulomb repulsion, the first term of $\mathcal{H}_{\text{corr}}$ in Eq. \ref{Hubbard} dominates to push the occupied $t_{2g}$ and $e_g$ spin majority states below the O-$p$ spectrum and to push the unoccupied spin minority states above in energy to stabilize the system in a CTI phase. The process of achieving CTI is shown through Fig. \ref{phase}(a-ii - a-iv). When the system is compressed, the octahedra are shrunk to yield a large $\Delta$ separating the $t_{2g}$ and $e_g$ (see Fig. \ref{phase}(a-vi). Simultaneously, with a reduction in the bond length, the kinetic hopping $t$ increases to reduce the effective Hund's coupling. Therefore, the competing $\Delta$ and $J_{\text{eff}}$ determine the spin occupancy of $t_{2g}$ and $e_g$ states. For $\Delta$ larger than $J_{\text{eff}}$, which occurs beyond a critical pressure $P_{c1}$, the $e_g$ states remain unoccupied and do not contribute to $\mathcal{H}_{\text{mag}}$ and $\mathcal{H}_{\text{corr}}$. This leads to a LS state ($S=1/2$; $t_{2g} ^{3 \uparrow,  2 \downarrow} e_g^0$). In the absence of strong correlation, the partially occupied $t_{2g}\downarrow$ sub-band yields a metallic state as shown in Fig. \ref{phase}(a-vii). With the inclusion of $\mathcal{H}_{\text{corr}}$, the strength of the first term, involving the Coulomb repulsion of the opposite spins, is effectively reduced to one-third ($=\langle n_{i\mu \uparrow} \rangle - \langle n_{i\mu \downarrow} \rangle $), and as a consequence, the $t_{2g}$ spin majority bands lie above the O-$p$ spectrum. Hence, CTI ceases to exist. However, the second term, involving the inter-orbital Coulomb repulsion, separates the $t_{2g}$ spin minority bands into two subbands and creates a Mott gap at the Fermi (see Fig. \ref{phase}a-viii). The Mott gap increases with an increase in $U$ (see Fig. \ref{phase}(a-ix)). From our first-principles calculation, we found that $P_{c1}$ increases with $U$, which is also reflected in the local spin moment (LSM) plot in Fig. \ref{phase}(a-x).

To provide a quantitative explanation of the stability as well as transition of the phases, in Fig. \ref{phase}(c) and (d), we have plotted the relative energies as a function of volume for $U=3$ and $5$ eV, respectively.
For the pressure-free system, the $P2_1nm$ HS phase (green curve) has lower energy as this curve lies distinctly below the blue curve indicating the $Pnnm$ HS phase (blue curve). The latter phase stabilizes over the former beyond $P_{c2}$. However, with further application of pressure, the energy difference between HS and LS (magenta curve) reduces, and the latter wins over the former at a critical volume $P_{c1}$. Unlike $P_{c1}$, $P_{c2}$ ($\approx 45$ GPa) is less sensitive to the Hubbard parameter $U$ as can be seen in Fig. \ref{phase}(a). The value of 45 GPa is in reasonable agreement with the one reported by Gleason \textit{et al.} (43 GPa) \cite{gleason_2013} and Insixiengmay \textit{et al.} (37 GPa) \cite{theory_Insixiengmay}. However, the experimental studies predict a second-order phase transition at 18 GPa and a first-order phase transition between 45 - 60 GPa \cite{thompson_2020}. The former is attributed to hydrogen bond symmetrization, and the latter is attributed to possible HS-LS spin crossover. If we go by the observation of Ikeda \textit{et al.} \cite{FeOOH-JACS}, at 18 GPa, even though the system makes a transition to the $Pnnm$ phase, the hydrogen bond symmetrization is not complete, as the H atom still has the fractional occupancy at two different Wyckoff positions, which escapes the scrutiny by DFT methods designed to study the periodic systems. However, we find that at $P_{c2}$, a perfect hydrogen bond symmetrization occurs in agreement with both theory and experiment.\\

\begin{figure}[hbt!]
    \centering
   \includegraphics[width=1\linewidth]{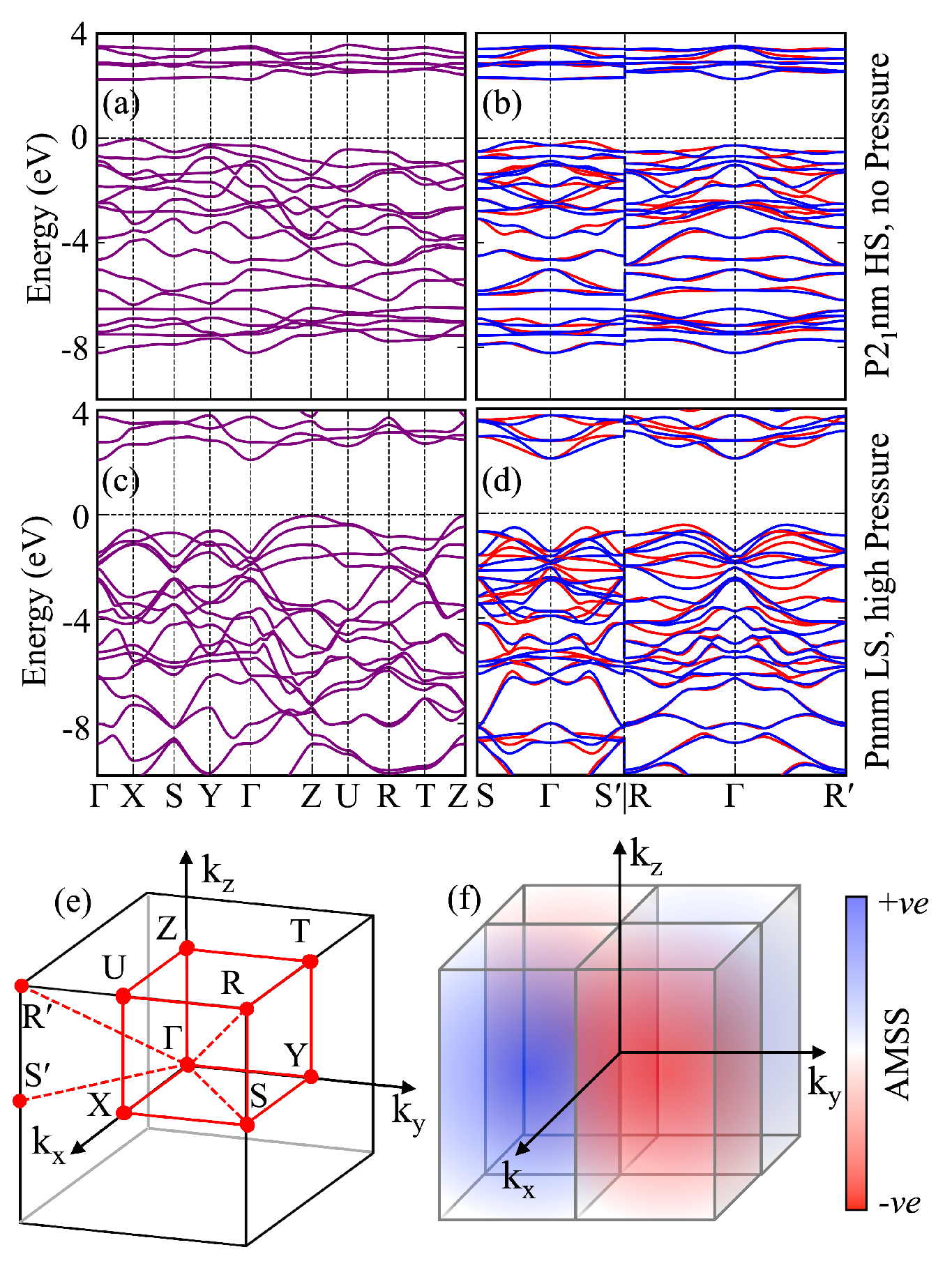}
    \caption{\textbf{Spin polarised band structure ($\boldsymbol{U = 5}$ eV) revealing altermagnetic characteristics of $\boldsymbol{\epsilon}$-FeOOH.} (a-b) the band structure, in the non-centrosymmetric $P2_1nm$ phase and at zero pressure, along the nodal plane showing antiferromagnetic band degeneracy (a) and away from the nodal plane showing AMSS (b). (c-d) same as the first row, but for the centrosymmetric $Pnnm$ phase at high pressure. (e) The Brillouin zone showing the high symmetric $k$-points. (f) Demonstration of nodal planes ($k_x-k_z$ and $k_y-k_z$) and anti-nodal zones (blue and red). Higher the intensity, higher the strength of the AMSS width. As we move from blue to red across a nodal plane, the sign of spin-split reverses.}
    \label{band}
\end{figure}

\noindent
\textbf{Altermagnetism and its characterization}

\noindent
Stabilization $A$-type N\'eel ordering in $\epsilon$-FeOOH, encourages us to examine the symmetry compatibility of this compound to exhibit altermagnetism. In the $P2_1nm$ phase, the three symmetry operations: one two fold screw, one mirror, and one glide along with identity $\mathcal{E}$ form the crystallographic Laue group, i.e., $\boldsymbol{G} = \{\mathcal{E}, \mathcal{C}_{2x}t_{\frac{1}{2},0,\frac{1}{2}}, \mathcal{M}_z, \mathcal{M}_y t_{\frac{1}{2},0,\frac{1}{2}} \}$. The elements of $\boldsymbol{G}$ can be divided into two spin halving subgroups $\boldsymbol{H}_{31} = \{ \mathcal{E}, \mathcal{M}_z\}$ and $\boldsymbol{(G-H)} = \{ \mathcal{C}_{2x}t_{\frac{1}{2},0,\frac{1}{2}}, \mathcal{M}_y t_{\frac{1}{2},0,\frac{1}{2}} \}$ such that they form type III Laue spin group $\boldsymbol{R}_s = [\mathcal{E}||\boldsymbol{H}] + [\mathcal{C}_2 || \boldsymbol{(G-H)}]$. For the case of the $Pnnm$ phase, the hydrogen bond symmetrization, in addition to the existing symmetries, introduces two more symmetry operations, namely, inversion ($\mathcal{I}$) and proper rotation. They reconstruct the crystallographic Laue group
$\boldsymbol{G} = \{\mathcal{E}, \mathcal{I}, \mathcal{C}_{2x}t_{\frac{1}{2},\frac{1}{2},\frac{1}{2}}, \mathcal{C}_{2y}t_{\frac{1}{2},\frac{1}{2},\frac{1}{2}}, \mathcal{C}_{2z}, \mathcal{M}_x t_{\frac{1}{2},\frac{1}{2},\frac{1}{2}}, \mathcal{M}_y t_{\frac{1}{2},\frac{1}{2},\frac{1}{2}}, \mathcal{M}_z \}$, where 
$\boldsymbol{H} = \{\mathcal{E}, \mathcal{I}, \mathcal{C}_{2z}, \mathcal{M}_z \}$ and

\noindent
$\boldsymbol{(G -H)} = \{\mathcal{C}_{2x}t_{\frac{1}{2},\frac{1}{2},\frac{1}{2}}, \mathcal{C}_{2y}t_{\frac{1}{2},\frac{1}{2},\frac{1}{2}}, \mathcal{M}_x t_{\frac{1}{2},\frac{1}{2},\frac{1}{2}}, \mathcal{M}_y t_{\frac{1}{2},\frac{1}{2},\frac{1}{2}} \}$, 

\noindent
A compound lying in the type III Laue spin group exhibits altermagnetism if there exist no inversion or translation symmetries in the halving subgroup $\boldsymbol{G-H}$, but there must exist a proper or improper rotational symmetry $A$ ($\in \boldsymbol{G-H}$) such that $\boldsymbol{G-H} = A \boldsymbol{H}$. As it is the case for both $P2_1nm$ and $Pnnm$ phases, $\epsilon$-FeOOH is an altermagnet in both phases.

The $Pnnm$ phase of $\epsilon$-FeOOH is centrosymmetric, and the inversion symmetry connects the same spin sublattices as it belongs to the halving subgroup $\boldsymbol{H}$. As a result, the antiferromagnetic spin degeneracy in the band structure is lifted to induce altermagnetism. The presence of $[\mathcal{T}_{\sigma}||\mathcal{M}_x]$ and $[\mathcal{T}_{\sigma}||\mathcal{M}_y]$ construct the $k_x-k_z$ and $k_y-k_z$ planes as nodal planes (see appendix; Eq. D7 and D8), thereby yielding a centrosymmetric $d$-wave altermagnet. Furthermore, since the $\epsilon$-FeOOH manifests collinear magnetism, in the non-relativistic limit, even though the system is noncentrosymmetric as in the $P2_1nm$ phase, the bands undergo $[ \mathcal{E}_{\sigma} || \mathcal{T}_L ]$ operation (identity operation on spin space and time-reversal operation on lattice space), which ensures AMSS $\Delta E(\boldsymbol{k}) = E(\sigma,\boldsymbol{k}) - E(-\sigma,\boldsymbol{k}) = \Delta E(-\boldsymbol{k})$ \cite{libor1, Arindom}. Interestingly, in the $P2_1nm$ phase, $[\mathcal{T}_{\sigma}||\mathcal{M}_y]$, along with $[ \mathcal{E}_{\sigma} || \mathcal{T}_L ]$, again construct the $k_x-k_z$ and $k_y-k_z$ planes nodal planes (see appendix; Eq. D10 and D12) to yield a rare noncentrosymmetric $d$-wave altermagnet. The other space groups that exhibits 3D collinear noncentrosymmetric altermagnets reported so far are $I4_1md$ (GdAlSi \cite{GdAlSi}, GdAlGe \cite{GdAlGe}), $P\Bar{6}2c$ (FeS \cite{FeS}), $P\bar{6}m2$ (doped CrSb \cite{Arindom}), and $P2_1$ (RbMnPO$_4$ \cite{Arindom}).

We will now validate the existence of altermagnetism through band structure, which is shown in Fig. \ref{band}, and examine the impact of several phase transitions on altermagnetism. The upper (Figs. \ref{band}(a,b)) and middle panels (Figs. \ref{band}(c,d)) show the band structure for the HS phase with the $P2_1nm$ space group and the LS phase with the $Pnnm$ space group, respectively. The band structure on the left demonstrates antiferromagnetic band degeneracy as the $k$-path considered lies on the nodal plane. The band structures on the right show the AMSS to reflect the altermagnetism in both HS and LS phases.

When we compare the band structure of HS and LS phases, we notice two striking differences. Firstly, the valence band maximum (VBM) and conduction band minimum (CBM) shift their position. In the case of LS, the CBM lies at $\Gamma$ with a well-dispersive parabolic band (see Fig. \ref{band}(c)), and the VBM lies at $Z$, which also has large dispersion. In the case of HS, the bands are weakly dispersed. There is a near absence of CBM as the conduction band edge is nearly flat across $\Gamma-X-S-Y$ (see Fig. \ref{band}(a)), while the VBM shifts from $Z$ to $X$. The alteration in the behavior of VBM and CBM can have consequences on the opto-electronics behavior of the system.

Secondly, the altermagnetic spin splitting (AMSS) is stronger for LS and weaker (almost reduced by half) for HS. This can be explained as follows. For the LS, the reduction in the bond length enhances the covalent hybridization, and therefore, the bands are expected to be more dispersed. On top of it, since the LS phase is MI, the spin carrying Fe-$d$ bands occupy the Fermi, while the O-$p$ bands lie below, which is substantiated through the atom resolved DOS (see Fig. \ref{phase}(b)). The spin-carrying bands are expected to show larger AMSS as compared to the O-$p$ bands, which are at best spin-induced due to hybridization. As the HS phase is a CTI, the localized Fe-$d$ bands lie far away from the Fermi energy $E_F$, while the spin-induced O-$p$ bands lie close to the $E_F$. These weakly spin-polarized bands generate weak AMSS.\\

\begin{figure}[hbt!]
    \centering
   \includegraphics[width=1\linewidth]{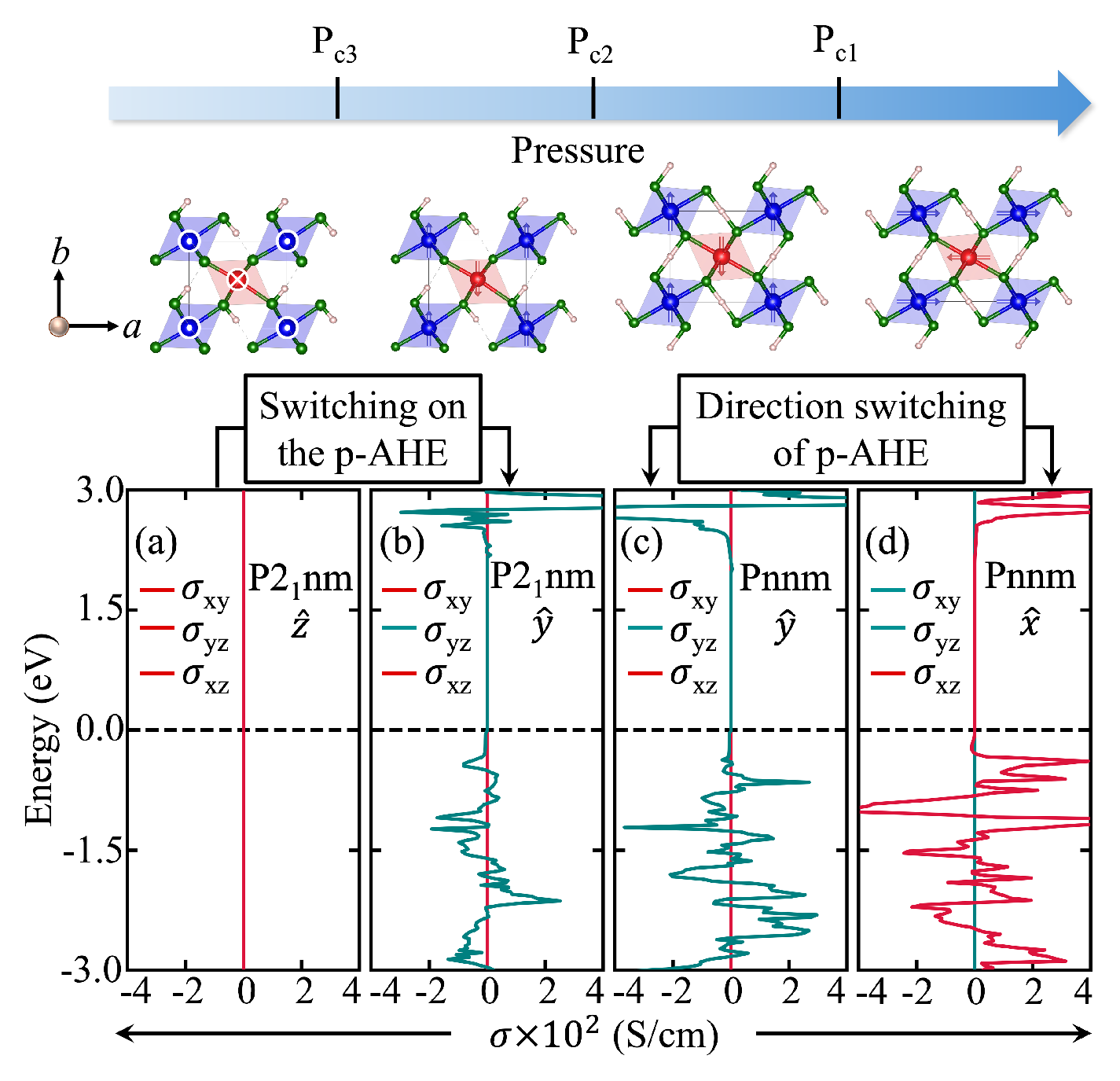}
    \caption{\textbf{Piezo-anomalous Hall effect in altermagnetic $\boldsymbol{\epsilon}$-FeOOH.} AHC for the structure (a) below $P_{c3}$, (b) between $P_{c3}$ and $P_{c2}$, (c) between $P_{c2}$ and $P_{c1}$, and (d) above $P_{c1}$. The critical pressures ($P_c$s) are indicated in Fig. \ref{toc}. We observe finite AHC when the spin quantization axis is $\hat{y}$ and $\hat{x}$. As the spin quantization axis switches at $P_{c3}$ and $P_{c1}$, the system is an ideal altermagnetic material to exhibit piezo-anomalous Hall effect (p-AHE).}
    \label{ahe}
\end{figure}

\noindent
\textbf{Magnetic anisotropy}

\noindent
Concerning the quantum transport, one of the expectations to make altermagnetic materials useful for spintronics applications is to demonstrate anomalous Hall conductivity (AHC) in them. The basic criterion here is to see that they have a preferred spin quantization axis that is symmetry compatible to exhibit AHC. For $\epsilon$-FeOOH, if the preferred spin orientation is along $\hat{x}$ or $\hat{y}$, then AHC is expected. As we will demonstrate here, the applied pressure, in addition to the aforementioned structural and magnetic phase transition, alters the spin quantization axis.

For the equilibrium configuration, the $\hat{x}$ axis remains as hard axis of the spin orientation, while both $\hat{y}$ and $\hat{z}$ axes become energetically favorable as energy difference of $\Delta E_{yz} \ (= E_y - E_z)$ tends to zero ($\approx -0.05$ meV). If the external pressure is applied or the correlation strength is reduced, the $\hat{y}$ quantization becomes distinctly energetically favorable beyond the critical pressure $P_{c3}$, as shown in Fig. \ref{phase}(a), with $\Delta E_{yz}$ reaching its peak value of $-0.3$ meV. Aligning with the above observation, the recent experiment by Ikeda \textit{et al.} reported the spin quantization axis as $\hat{z}$ for the equilibrium configuration and switches to $\hat{y}$ with pressure \cite{FeOOH-JACS}. Interestingly, at the critical pressure $P_{c1}$, where the HS-LS spin crossover occurs, $\hat{x}$ quantization becomes favorable with $\Delta E_{yx}$ more than $2$ meV. This observation opens up scopes for the experimenter to validate and establish spin quantization switching.\\

\noindent
\textbf{Direction-dependent piezo-anomalous Hall effect} 

\noindent
The role of spin quantization on the anomalous Hall transport in the $\epsilon$-FeOOH is discussed next. This is premised on the fact that the existence of AHC and the allowed components of its tensor depend on the N\'eel vector orientation and the corresponding magnetic space group (MSG). The symmetry-allowed AHC tensor components for different magnetic configurations are summarized in Tables III and IV of the appendix. For the $P2_1nm$ phase in equilibrium condition, the N\'eel vector is oriented along the $\hat{z}$ direction, as discussed earlier. The corresponding magnetic symmetry forbids all components of the AHC tensor, leading to a vanishing intrinsic AHC. Beyond the critical pressure $P_{c3}$, the N\'eel vector changes to the $\hat{y}$ direction, which in turn, changes the MSG. The new symmetry elements allow the $\sigma_{yz}$ component of the AHC, while all other tensor components remain forbidden. The same is the case for the $Pnnm$ phase with HS state and spin quantization along $\hat{y}$. Upon further application of pressure beyond $P_{c1}$, where the spin crossover occurs, the N\'eel vector reorients along the $\hat{x}$ axis, which further modifies the corresponding symmetries. In the LS state, only the $\sigma_{xz}$ component of AHC is allowed. In Fig. \ref{ahe}, we have presented the numerical results for various components of the AHC using Berry curvature in different situations allowed by the corresponding symmetries. The magnitude and direction switching of AHC with pressure are clearly seen in the figure. Furthermore, from the figure, we observe that the magnitude of the AHC increases with pressure. The details of the numerical calculation are provided in the appendix. Therefore, $\epsilon$-FeOOH is an ideal compound to establish piezo-anomalous Hall response, where AHC can be switched on and off or its direction can be changed by tuning the pressure.

\section*{Discussion}

Necessitated by experimentally observed pressure-tunable electronic and magnetic structures in $\epsilon$-FeOOH, in this work, we have established a pressure-correlation phase diagram for it. Through this phase diagram, we reaffirmed the existence of two critical pressures driving the structural transition ($P_{c2}$) and spin crossover ($P_{c1}$). While $P_{c2}$ is the least influenced by the correlation, $P_{c1}$ is sensitive to it. In addition, we identified a third critical pressure $P_{c3}$ across which the spin quantization axis changes from $\hat{z}$ to $\hat{y}$, which is broadly consistent with the very recent experimental observation by Ikeda \textit{et al.} \cite{FeOOH-JACS}. $P_{c3}$ is found to be highly dependent on the electron-electron correlation. We established another change in the spin quantization axis (from $\hat{y}$ to $\hat{x}$) at $P_{c1}$ that was missing in earlier studies. The present work, through a thorough analysis of electronic structure and mapping that to a multiorbital Hubbard Hamiltonian, provides the underlying mechanisms governing the mechanism of HS and LS states and their transitions. It demonstrates that due to competing the crystal field and effective Hund's coupling, the HS state is synonymous with the CTI electronic state, and the LS state is synonymous with the MI electronic state. Our group theoretical analysis, along with the band structure studies, infer that the compound is altermagnetic in both $P2_1nm$ (noncentrosymmetric) and $Pnnm$ (centrosymmetric) phases. Therefore, $\epsilon$-FeOOH is a unique system to exhibit pressure switchable centrosymmetric to non-centrosymmetric altermagnet. Furthermore, we find that the system exhibits finite AHC provided the spin quantization axis is either $\hat{y}$ or $\hat{x}$. Therefore, the critical pressures $P_{c1}$ and $P_{c3}$ imply the realization of piezo-anomalous Hall effect, which is promising for spintronics applications.

\section*{Methods}

In this study, DFT calculations are performed using the plane-wave projector augmented-wave (PAW) method \cite{paw1, paw2}, as implemented in the Vienna \textit{Ab initio} Simulation Package (VASP) \cite{vasp}, to investigate the electronic and magnetic properties of the compound. The Perdew-Burke-Ernzerhof generalized gradient approximation is chosen for the exchange correlation functional. The Brillouin zone is sampled using a $\Gamma$-centered $6 \times 6 \times 8$ $k$-point mesh, which is decided from a convergence test. The plane-wave energy cutoff of 600 eV is adopted. To account for the effects of electron correlations, the DFT+U approach is employed within the rotationally invariant formalism proposed by Dudarev \cite{dudarev}, using the effective Hubbard parameter $U_{\text{eff}}=U-J$. To evaluate the AHC, a tight-binding Hamiltonian was constructed using maximally localized Wannier functions constructed using the Wannier90 code \cite{mostofi2014updated}. The AHC was subsequently calculated from this Wannier-based Hamiltonian using the WannierBerri code \cite{tsirkin2021high}.

\section*{Acknowledgement}
AM thanks MoE India for the PMRF fellowship. The authors acknowledge IIT Madras for the computational resources. The authors acknowledge Babuji Dandigunta for useful discussions.

\appendix
\section{Structural details of $\epsilon$-FeOOH}

\begin{figure}[hbt!]
    \centering
   \includegraphics[width=1.0\linewidth]{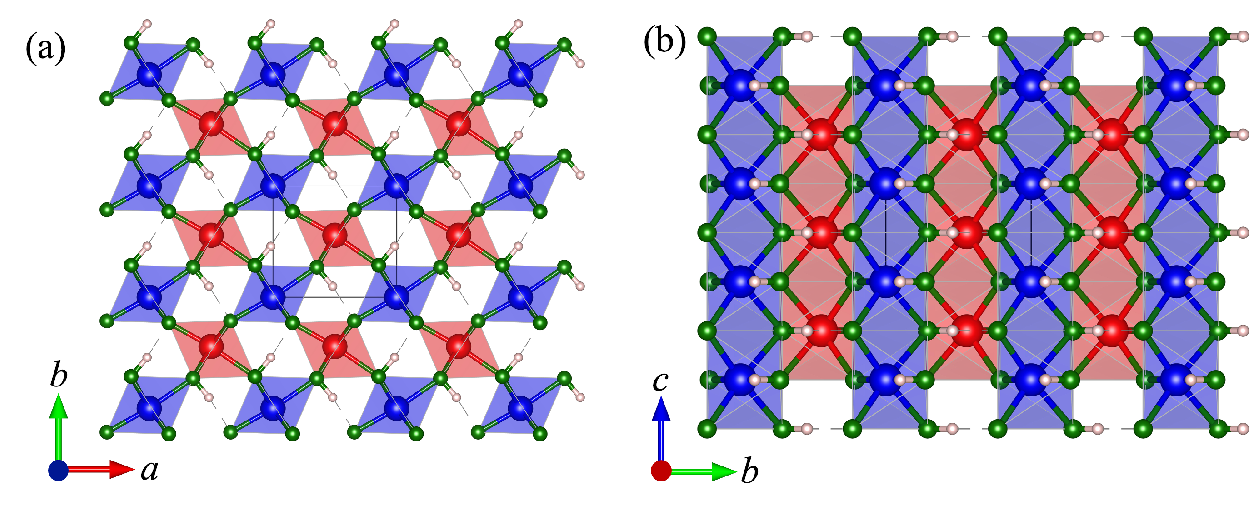}
    \caption{\textbf{Crystal structure of $\boldsymbol{\epsilon}$-FeOOH in $\boldsymbol{P2_1nm}$ phase.} (a) and (b) show the view from $c$ and $a$ axes, respectively.}
    \label{crystal}
\end{figure}

The $\epsilon$ phase of FeOOH, a high-pressure quenchable polymorph of its $\alpha$ phase (goethite), was first synthesized by Chenavas \textit{et al.} in 1973, and has an InOOH-type crystal structure with $P2_1nm$ space group \cite{FeOOH-synthesis, FeOOH-XRD}. 
As shown in Fig. \ref{crystal}, each Fe atom is surrounded by six oxygen atoms forming a FeO$_6$ stretched octahedron. The octahedra are edge-shared along $c$. Along $a$, the octahedra are separated each other through a void. Furthermore, the octahedra are layer stacked along $b$, and the stretched axis of the octahedra along the diagonal of $ab$ plane are zig-zagged to constitute a cooperative distortion. 

Oxygen from every alternate layer forms hydroxy bonds, while oxygen from the other alternate layers forms hydrogen bonds in such a way that there is an O$-$H$\cdots$O bond lying diagonally in the void. The lattice parameters of the unit cell are optimized to be $a=4.99$ \AA, $b=4.45$ \AA, $c=3.02$ \AA, which are in close agreement with experimental reported values of $a=4.93918$ \AA, $b=4.43693$ \AA, $c=2.99433$ \AA \cite{FeOOH-JACS}.

Upon applying hydrostatic pressure to this compound, a structural phase transition occurs. The $P2_1nm$ phase changes to the $Pnnm$ phase, which is further verified by estimating the relative energy difference obtained from the \textit{ab-initio} calculations.
The structural transition from $P2_1nm$ to $Pnnm$ phase is concurrent with the vanishing of the hydrogen bond (i.e., H$\cdots$O). With an intermediate pressure ($\approx 18$ GPa) \cite{FeOOH-JACS}, the O$-$H$\cdots$O is now replaced with a single O$-$H bond. However, there is a $50\%$ probability that the H atom is attached to the lower oxygen layer and another $50\%$ probability is that it is connected to the upper oxygen layer. As the pressure increases, the O$-$H$\cdots$O bond becomes an O$-$H$-$O bond, where H is rigidly connected to both the oxygen atoms instead of flipping from one to the other. As the DFT calculations are done on a static equilibrium configuration, the fractional occupancies cannot be accommodated, and we have considered the O$-$H$-$O bond both for the intermediate and high-pressure $Pnnm$ phase.

\section{Pressure vs Volume curve}

\begin{figure}[hbt!]
    \centering
   \includegraphics[width=1.0\linewidth]{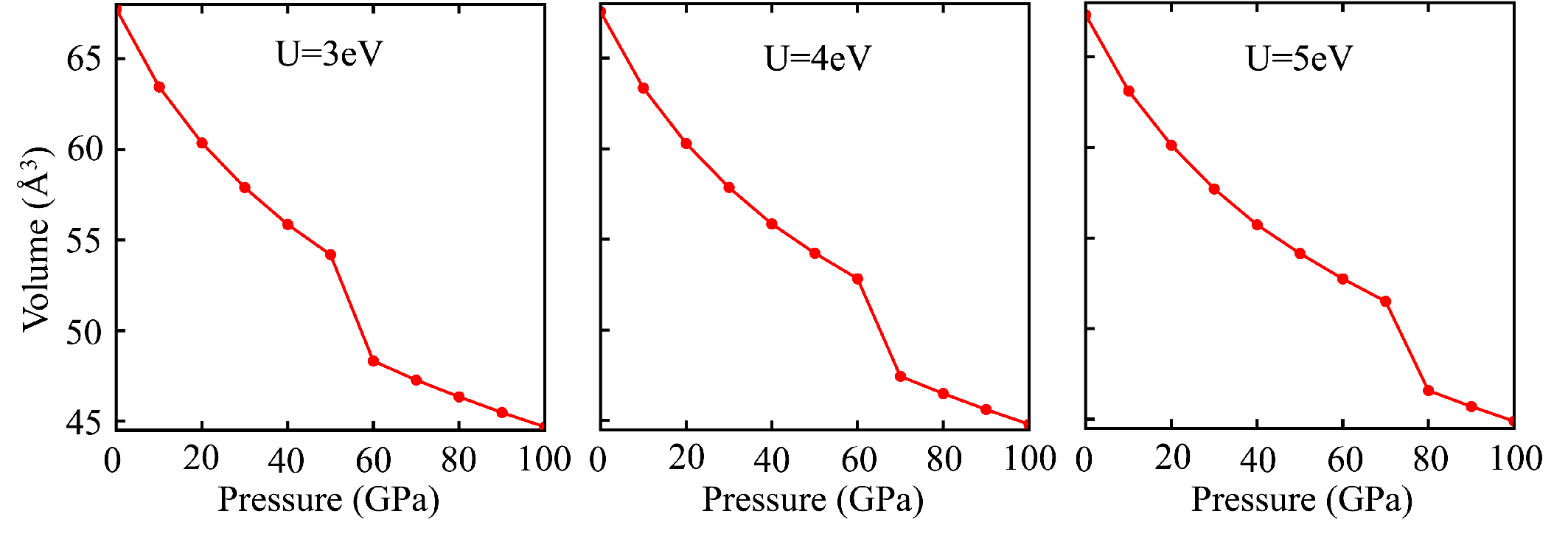}
    \caption{\textbf{Unit cell volume as a function of pressure for different correlation strengths.} The volume shrunk suddenly when the system undergoes a transition from HS state to LS state, and the pressure at which this discontinuity occurs increases with increases in correlation strength.}
    \label{pressure_volume}
\end{figure}

The volume of the unit cell for a given pressure and correlation strength is obtained by relaxing the atomic positions and cell volume in DFT calculations. The results are shown in Fig. \ref{pressure_volume}. From the figure, we see a sudden reduction in volume at the critical pressure $P_{c1}$, where HS-LS crossover occurs. This discontinuity was reported in Ref. \cite{thompson_2020} as a first-order phase transition. Furthermore, we observe that $P_{c1}$ increases with $U$, which suggests that HS-LS crossover not only occurs due to an increase in the crystal field as reported by Insixiengmay \textit{et al.} \cite{theory_Insixiengmay}, but also that a competition among the crystal field, Hund's coupling, and electron correlation is involved in it.

\section{Multi-orbital Hubbard Model}

The multi-orbital Hubbard Hamiltonian is written as
\begin{widetext}
\begin{eqnarray}
    \mathcal{H} &=& \sum_{i,\mu} \epsilon_{i\mu} c_{i\mu}^{\dagger}c_{i\mu} + \sum_{i,j,\mu,\nu} t_{i\mu j \nu} c_{i\mu}^{\dagger} c_{j\nu} + \text{H.c.} + \mathcal{H}_{\text{corr}}  + \mathcal{H}_{\text{mag}}\\
    \mathcal{H}_{\text{corr}} &=& U\sum_{i,\mu} n_{i\mu \uparrow} n_{i\mu \downarrow} + \left( U'-\frac{J}{2} \right)\sum_{i,\mu < \nu} n_{i\mu} n_{i\nu} = \mathcal{H}_1 + \mathcal{H}_2 \\
    \mathcal{H}_{\text{mag}} &=& - 2J\sum_{i,\mu< \nu} S_{i\mu}^z \cdot S_{i\nu}^z
\end{eqnarray}
where $U' = U - 2J$, and,
\begin{eqnarray}
    \mathcal{H}_1 &=& U\sum_{i,\mu} n_{i\mu \uparrow} n_{i\mu \downarrow} = U \sum_{i,\mu} (n_{i\mu \uparrow} \langle n_{i\mu \downarrow} \rangle + n_{i\mu \downarrow} \langle n_{i\mu \uparrow} \rangle - \langle n_{i\mu \downarrow} \rangle \langle n_{i\mu \uparrow} \rangle)\\
    \mathcal{H}_2 &=& \left( U'-\frac{J}{2} \right)\sum_{i,\mu < \nu} n_{i\mu} n_{i\nu} = \left( U'-\frac{J}{2} \right) \sum_{i,\mu < \nu} (n_{i\mu} \langle n_{i\nu} \rangle + n_{i\nu} \langle n_{i\mu} \rangle - \langle n_{i\mu} \rangle \langle n_{i\nu} \rangle)\\
    \mathcal{H}_{\text{mag}} &=& -2J\sum_{i,\mu< \nu} S_{i\mu}^z \cdot S_{i\nu}^z = -\frac{J}{2} \sum_{i,\mu < \nu} (n_{i\mu \uparrow} (\langle n_{i\nu \uparrow} \rangle - \langle n_{i\nu \downarrow} \rangle) + n_{i\nu \uparrow} (\langle n_{i\mu \uparrow} \rangle - \langle n_{i\mu \downarrow} \rangle) \nonumber \\
    && - n_{i\mu \downarrow} (\langle n_{i\nu \uparrow} \rangle - \langle n_{i\nu \downarrow} \rangle) - n_{i\nu \downarrow} (\langle n_{i\mu \uparrow} \rangle - \langle n_{i\mu \downarrow} \rangle)  -\langle n_{i\mu \uparrow} \rangle \langle n_{i\nu \uparrow} \rangle + \langle n_{i\mu \uparrow} \rangle \langle n_{i\nu \downarrow} \rangle + \langle n_{i\mu \downarrow} \rangle \langle n_{i\nu \uparrow} \rangle - \langle n_{i\mu \downarrow} \rangle \langle n_{i\nu \downarrow} \rangle) \nonumber\\
\end{eqnarray}
\end{widetext}

\begin{figure}[hbt!]
    \centering
   \includegraphics[width=1\linewidth]{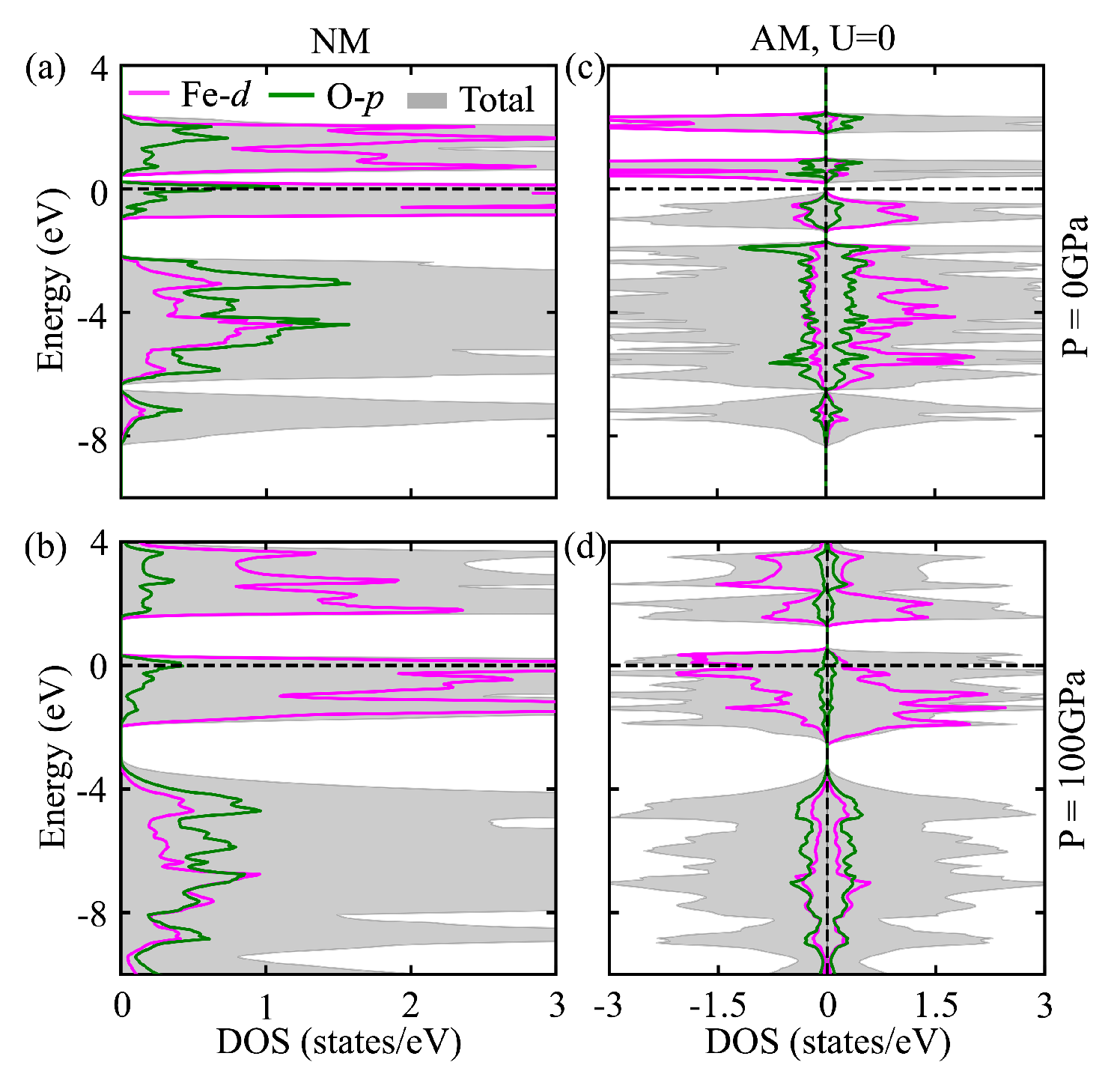}
    \caption{\textbf{Density of states of Fe in $\epsilon$-FeOOH with spin up majority.} (a) and (b) represent the nonmagnetic DOS at P=0 and 100 GPa, respectively, while (c) and (d) represent the altermagnetic DOS for the same with $U=0$ eV.}
    \label{dos_hypo}
\end{figure}

Since $\epsilon$-FeOOH stabilizes in the $A$-type antiferromagnetic spin ordering, the two Fe atoms exhibit opposite spins. Here, we analyze the occupation of different orbitals of the Fe atom, which has the spin-up majority. The same analysis applies to the other Fe atom. The electron occupation numbers for the HS and LS states are tabulated in Table \ref{occ_table}. From the table and the above-mentioned equations, we see that in the HS state, the Hamiltonian $\mathcal{H}_2$ does not create an energy gap between the spin-up and spin-down states. For $U = 0$, Hamiltonian $\mathcal{H}_1$ also has no role in the formation of the energy gap. Therefore, the Hund's coupling energy can be estimated from Hamiltonian $\mathcal{H}_3$. The energy gap for $e_g$ orbitals from the DFT-produced DOS is obtained as $J_{\text{eff}} = 3$ eV. Hence, $3 = -\frac{J}{2}(1+1)$ (see Fig. \ref{pressure_volume}). From here, we can estimate $J \approx 3$ eV.

If the external pressure is applied to it, the volume of the unit cell decreases, which results in an increase in crystal field splitting $\Delta$. At the moment $\Delta$ exceeds the $J$, the system jumps from the HS state to the LS state. From the nonmagnetic DOS, we find $\Delta \approx 4$ eV at 100 GPa pressure. Furthermore, the gap due to Hund's coupling here should be approximately $3 \times (2/3) = 2$ eV, which matches with the DFT obtained DOS with $U=0$ eV (see Fig. \ref{pressure_volume}).

\begin{table*}[hbt!]
\begin{center}
\caption{Electron occupation for the up spin and down spin $d$ orbitals in HS and LS state}
\setlength{\tabcolsep}{6pt}
\renewcommand{\arraystretch}{2}
\begin{tabular}{c| c c c c c | c c c c c}
\hline
\hline
Spin state & $d_1 \uparrow$ & $d_2 \uparrow$ & $d_3 \uparrow$ & $d_4 \uparrow$ & $d_5 \uparrow$ & $d_1 \downarrow$ & $d_2 \downarrow$ & $d_3 \downarrow$ & $d_4 \downarrow$ & $d_5 \downarrow$ \\
\hline
HS & $1$ & $1$ & $1$ & $1$ & $1$ & $0$ & $0$ & $0$ & $0$ & $0$ \\
LS &  $1$ & $1$ & $1$ & $0$ & $0$ & $2/3$ & $2/3$ & $2/3$ & $0$ & $0$ \\
\hline
\hline
\end{tabular}
\label{occ_table}
\end{center}
\end{table*}

\section{Altermagnetic symmetry and nodal planes}

For a collinear magnet, as is the case of $\epsilon$-FeOOH, under the non-relativistic limit the bands always have the $[ \mathcal{E}_{\sigma} || \mathcal{T}_L ]$ symmetry (where, $\mathcal{E}_{\sigma}$ is identity operation on spin space while $\mathcal{T}_L$ is the time-reveral operation on lattice space), which mimics the effect of the inversion symmetry in the band structure \cite{libor1, Arindom}, i.e.,
\begin{equation}
    [ \mathcal{E}_{\sigma} || \mathcal{T}_L ]E(k_x, k_y, k_z,\sigma) = E(-k_x, -k_y, -k_z,\sigma)
    \label{eq:TL}
\end{equation}
Therefore, the nonrelativistic bands of collinear spin systems always have the inversion symmetry irrespective of their centrosymmetricity. This ensures $\Delta E(\boldsymbol{k}) = \Delta E(-\boldsymbol{k})$, where $\Delta E = E(\sigma) - E(-\sigma)$.

In the centrosymmetric $Pnnm$ phase of $\epsilon$-FeOOH, the crystal symmetries impose the following constraints on the band structure.

\begin{eqnarray}
    \label{eq:I1}
    [\mathcal{E}_{\sigma} || \mathcal{I}] E(k_x, k_y, k_z, \sigma) &=& E(-k_x, -k_y, -k_z, \sigma)\\
    \label{eq:C2z1}
    [\mathcal{E}_{\sigma} || \mathcal{C}_{2z}] E(k_x, k_y, k_z, \sigma) &=& E(-k_x, -k_y, k_z, \sigma)\\
    \label{eq:Mz1}
    [\mathcal{E}_{\sigma} || \mathcal{M}_z] E(k_x, k_y, k_z, \sigma) &=& E(k_x, k_y, -k_z, \sigma)\\
    \label{eq:C2x1}
    [\mathcal{T}_{\sigma} || \mathcal{C}_{2x}] E(k_x, k_y, k_z, \sigma) &=& E(k_x, -k_y, -k_z, -\sigma)\\
    \label{eq:C2y1}
    [\mathcal{T}_{\sigma} || \mathcal{C}_{2y}] E(k_x, k_y, k_z, \sigma) &=& E(-k_x, k_y, -k_z, -\sigma)\\
    \label{eq:Mx1}
    [\mathcal{T}_{\sigma} || \mathcal{M}_x] E(k_x, k_y, k_z, \sigma) &=& E(-k_x, k_y, k_z, -\sigma)\\
    \label{eq:My1}
    [\mathcal{T}_{\sigma} || \mathcal{M}_y] E(k_x, k_y, k_z, \sigma) &=& E(k_x, -k_y, k_z, -\sigma)
\end{eqnarray}

Here we see that the results in Eqs. \ref{eq:TL} and \ref{eq:I1} are identical as discussed earlier.  $[\mathcal{T}_{\sigma}||\mathcal{M}_x]$ and $[\mathcal{T}_{\sigma}||\mathcal{M}_y]$ make the $k_x-k_z$ and $k_y-k_z$ plane spin degenerate, respectively, while $[\mathcal{T}_{\sigma}||\mathcal{C}_{2x}]$ and $[\mathcal{T}_{\sigma}||\mathcal{C}_{2y}]$ make the $k_x$ and $k_y$ axis spin degenerate, respectively. This makes the system a centrosymmetric $d$ wave altermagnet as shown in Fig. 3(a) in the main text.

In the $P2_1nm$ phase, the crystal symmetry-driven constraints on the band structure are given by,
\begin{eqnarray}
     \label{eq:Mz}
    [\mathcal{E}_{\sigma} || \mathcal{M}_z] E(k_x, k_y, k_z, \sigma) &=& E(k_x, k_y, -k_z, \sigma)\\
    \label{eq:My}
    [\mathcal{T}_{\sigma} || \mathcal{M}_y] E(k_x, k_y, k_z, \sigma) &=& E(k_x, -k_y, k_z, -\sigma)\\
    \label{eq:C2x}
    [\mathcal{T}_{\sigma} || \mathcal{C}_{2x}] E(k_x, k_y, k_z, \sigma) &=& E(k_x, -k_y, -k_z, -\sigma)
\end{eqnarray}
Note that inversion symmetry is broken in this compound. Here, only $[\mathcal{T}_{\sigma}||\mathcal{M}_y]$ and $[\mathcal{T}_{\sigma}||\mathcal{C}_{2x}]$ can establish a relation between the band energies of two opposite spins, as evident from Eqs. \ref{eq:My} and \ref{eq:C2x}. Out of this two $[\mathcal{T}_{\sigma}||\mathcal{M}_y]$ makes the $k_x-k_z$ plane spin degenerate, while $[\mathcal{T}_{\sigma}||\mathcal{C}_{2x}]$ makes the $k_x$ axis spin degenerate. Moreover, the above-mentioned symmetries followed by $[ \mathcal{E}_{\sigma} || \mathcal{T}_L ]$ provide a similar effect of $[\mathcal{T}_{\sigma}||\mathcal{C}_{2y}]$ and $[\mathcal{T}_{\sigma}||\mathcal{M}_x]$, which results in a spin band degenerate $k_y$ axis and $k_y-k_z$ plane, respectively, as shown below.

\begin{eqnarray}
    [\mathcal{E}_{\sigma} || \mathcal{T}_L][\mathcal{T}_{\sigma} || \mathcal{M}_y] E(k_x, k_y, k_z, \sigma) &=& [\mathcal{T}_{\sigma} || \mathcal{M}_x] E(k_x, k_y, k_z, \sigma) \nonumber \\
    &=& E(-k_x, k_y, k_z, -\sigma) \nonumber \\
\end{eqnarray}

As a whole, the $P2_1nm$ phase of $\epsilon$-FeOOH also has two distinct spin band degenerate planes, namely $k_x-k_z$ and $k_y-k_z$ (the degenerate axes are included in the degenerate planes), which are known as the altermagnetic nodal plane. Therefore, the $P2_1nm$ phase is a noncentrosymmetric $d$ wave altermagnet. 

\section{Anomalous Hall effect}

\begin{table}
\begin{center}
\caption{AHC in $P2_1nm$ with $z$ spin (MSG: $Pmn2_1$ (No: 31.123)) and $y$ spin (MSG: $Pm'n'2_1$ (No: 31.127))}
\setlength{\tabcolsep}{6pt}
\renewcommand{\arraystretch}{2}
\begin{tabular}{c| c c c}
\hline
\hline
Symmetry & $\sigma_{xy}$ & $\sigma_{xz}$ & $\sigma_{yz}$ \\
\hline
\multicolumn{4}{c}{$P2_1nm$ with spin direction $\hat{z}$}\\
\hline
$M_z$ & $\checkmark$ & $\times$ & $\times$ \\
$M_y t_{\frac{1}{2},0,\frac{1}{2}}$ &  $\times$ & $\checkmark$ & $\times$ \\
$C_{2x} t_{\frac{1}{2},0,\frac{1}{2}}$ & $\times$ & $\times$ & $\checkmark$ \\
\hline
\multicolumn{4}{c}{$P2_1nm$ with spin direction $\hat{y}$}\\
\hline
$\tau M_z$ & $\times$ & $\checkmark$ & $\checkmark$ \\
$\tau M_y t_{\frac{1}{2},0,\frac{1}{2}}$ &  $\checkmark$ & $\times$ & $\checkmark$ \\
$C_{2x} t_{\frac{1}{2},0,\frac{1}{2}}$ & $\times$ & $\times$ & $\checkmark$ \\
\hline
\hline
\end{tabular}
\label{tab1}
\end{center}
\end{table}

\begin{table}
\begin{center}
\caption{AHC in $Pnnm$ with $x$ and $y$ spin (MSG: $Pnnm'$ (No: 58.398))}
\setlength{\tabcolsep}{6pt}
\renewcommand{\arraystretch}{2}
\begin{tabular}{c | c c c}
\hline
\hline
Symmetry & $\sigma_{xy}$ & $\sigma_{xz}$ & $\sigma_{yz}$ \\
\hline
\multicolumn{4}{c}{$Pnnm$ with spin direction $\hat{x}$}\\
\hline
$\tau C_{2x}t_{\frac{1}{2},\frac{1}{2},\frac{1}{2}}$ & $\checkmark$ & $\checkmark$ &$\times$  \\
$ C_{2y}t_{\frac{1}{2},\frac{1}{2},\frac{1}{2}}$ & $\times$ & $\checkmark$ &  $\times$\\
$\tau C_{2z}$ & $\times$ &$\checkmark$  &$\checkmark$  \\
$\tau M_{x}t_{\frac{1}{2},\frac{1}{2},\frac{1}{2}}$ & $\checkmark$ & $\checkmark$  &$\times$  \\
$M_{y}t_{\frac{1}{2},\frac{1}{2},\frac{1}{2}}$ & $\times$ & $\checkmark$ & $\times$ \\
$\tau M_{z}$ &$\times$  & $\checkmark$  & $\checkmark$  \\
\hline
\multicolumn{4}{c}{$Pnnm$ with spin direction $\hat{y}$}\\
\hline
$C_{2x}t_{\frac{1}{2},\frac{1}{2},\frac{1}{2}}$ & $\times$ & $\times$ &$\checkmark$  \\
$\tau C_{2y}t_{\frac{1}{2},\frac{1}{2},\frac{1}{2}}$ & $\checkmark$ & $\times$ &  $\checkmark$\\
$\tau C_{2z}$ & $\times$ &$\checkmark$  &$\checkmark$  \\
$M_{x}t_{\frac{1}{2},\frac{1}{2},\frac{1}{2}}$ & $\times$ & $\times$  &$\checkmark$  \\
$\tau M_{y}t_{\frac{1}{2},\frac{1}{2},\frac{1}{2}}$ & $\checkmark$ & $\times$ & $\checkmark$ \\
$\tau M_{z}$ &$\times$  & $\checkmark$  & $\checkmark$  \\
\hline
\hline
\end{tabular}
\label{tab2}
\end{center}
\end{table}

As FeOOH is an altermagnet whose magnetic orientation changes with pressure, we examined the possibility of pressure-induced AHE using magnetic symmetry analysis and first-principles calculations. The symmetry compatibility for the existence of finite AHC for different spin quantization axes is provided in Tables \ref{tab1} and \ref{tab2}. The Berry curvature, $\Omega_{n}^{\nu}(\mathbf{k})$, was calculated over the Brillouin zone using~\cite{gradhand2012first}
\begin{equation}
    \Omega_{n}^{\nu}(\textbf{k}) = -2 \, \text{Im} \sum_{n \neq {n'}} \frac{\langle n \vec{k}|v_{\gamma}({\vec{k})}|n'\vec{k}\rangle \langle n'\vec{k}|v_{\delta} (\vec{k}) |n\vec{k}\rangle}{\left(E_n (\vec{k}) - E_{n'} (\vec{k})\right)^2}.
\end{equation}
Here, $v_{\gamma}$ is the velocity operator, and $E_n(\mathbf{k})$ is the energy eigenvalue of the $n$th band. The AHC was then obtained using the Kubo formula~\cite{thouless1982quantized},
\begin{equation}
\sigma_{\gamma\delta}=\varepsilon_{\gamma\delta\nu}\frac{e^2}{\hbar}\sum_n\int_{\mathrm{BZ}}f(\epsilon_n(\mathbf{k}))\Omega_n^{\nu}(\mathbf{k})\frac{d\mathbf{k}}{(2\pi)^3},
\label{sigma_eqn}
\end{equation}
where $e$, $\hbar$, $\varepsilon_{\gamma\delta\nu}$, and $f(\epsilon_n(\mathbf{k}))$ denote the electron charge, reduced Planck's constant, the Levi-Civita symbol, and the Fermi--Dirac distribution function, respectively.
FeOOH crystallizes in the $P2_1nm$ and $Pnnm$ space groups in different pressure ranges. 

\bibliography{ref.bib}
\end{document}